\documentstyle[aaspp4]{article}
\begin{document}
 
\lefthead{Efremov, Elmegreen, \& Hodge}
\righthead{Relics of Gamma Ray Burst Explosions}
\slugcomment{IBM Log 94648; accepted by ApJ Letters}
 
\title{Giant Shells and 
Stellar Arcs as Relics of Gamma Ray Burst Explosions}

\author{Yuri N. Efremov\altaffilmark{1},
Bruce G. Elmegreen\altaffilmark{2}, Paul W.Hodge\altaffilmark{3}
}
\altaffiltext{1}{P.K. Sternberg Astronomical Institute, MSU,
Moscow 119899, Russia, efremov@sai.msu.su} 
\altaffiltext{2}{
IBM Research Division, T.J. Watson Research Center,
P.O. Box 218, Yorktown Heights, NY 10598, USA, bge@watson.ibm.com}
\altaffiltext{3}{Washington Astronomy Department, Box 351580, 
University of Washington, Seattle, WA 98195, USA,
hodge@astro.washington.edu}

\begin{abstract} Gamma Ray Burst (GRB) explosions 
are powerful and frequent enough to make kiloparsec-size
shells and holes in the interstellar media of spiral galaxies.
The observations of such remnants are summarized. 
Several observed shells contain no
obvious central star clusters and could be GRB remnants, but
sufficiently old clusters that could have formed them 
by supernovae and winds might be hard to detect. 
\end{abstract}
 
Subject headings: gamma rays: bursts --- galaxies: ISM --- 
ISM: bubbles --- ISM: supernova remnants

submitted ... 17 March 1998, accepted ... 15 May 1998

\section{Introduction}

The recent discovery of optical afterglows from gamma-ray bursters and
the realization that they are at cosmological distances and extremely
powerful, raises the issue of their interaction with the interstellar
medium of the host galaxy. This revives the old suggestion that some
large stellar and interstellar structures such as arcs of star clusters,
HI supershells, and dust rings, are caused by super-supernovae (e.g.,
Shklovsky 1960; Hayward 1964; Westerlund \& Mathewson 1966; Hodge 1967).

Optical counterparts to GRBs were found after 20 years of searching when
GRB970228 (Groot et al. 1997; van Paradijs et al. 1997) and GRB970508
(Bond 1997) were assigned accurate positions by the Xray satellite {\it
BeppoSAX} (Costa et al. 1997; Piro et al. 1998). Absorption lines in
GRB970508 (Metzger et al. 1997) place it at a redshift
of $z=0.8$ to 2.3, while the possible signature of extinction suggests
$z=1.09$ (Reichart 1998). There may also be faint galaxies around
GRB970228 (Sahu et al. 1997) and GRB970508 (Pedersen et al. 1998). 
For such distances, the gamma ray energy
alone is $\sim10^{51}$ ergs, and the total fireball energy can be
$10^{52}$ erg or more, considering the likely inefficiency of gamma
radiation (Waxman 1997; Rees \& M\'esz\'aros 1998). 
Optically, GRBs can outshine supernovae by a
factor of $\sim100$ (Pian et al. 1998; 
Paczy\'nski 1998), and the optical flux from the
afterglow can exceed the gamma ray and x-ray fluxes by the same factor
(Wijers, Rees \& M\'esz\'aros 1997).

GRBs and their afterglows at x-ray, optical, and radio wavelengths
presumably arise from synchrotron and inverse Comption radiation in the
shocked parts of relativistic fireballs and their surrounding
interstellar media (Paczy\'nski \& Rhoads 1993; M\'esz\'aros \& Rees
1997; Vietri 1997; Waxman 1997; Sari 1997). The energy could come from
the release of gravitational binding energy ($\sim10^{54}$ ergs) during
the rapid formation of a black-hole. This may occur for neutron stars
that acquire too much mass to be stable during binary coalescence
(Blinnikov et al. 1984) or Roche-lobe overflow from evolving companion
stars (Qin et al. 1998), or it may occur in ``failed supernovae''
(Woosley 1993) or ``hypernovae'' collapse of spinning massive stars
(Paczy\'nski 1998). The energy liberated in each of these events can be
much larger than the observed gamma ray and afterglow energies, so there
is a good possibility that a large amount of kinetic energy ($>10^{52}$
erg) in the form of expanding motions and hot gas remains to affect the
surrounding interstellar gas for several million years following the
explosion.

A possible connection between hypernovae events and HI supershells
without central star clusters was mentioned by Blinnikov \& Postnov
(1998) but not discussed in any detail. If the frequency of such events
is comparable to or higher than the frequency of neutron-star 
mergers or binary accretions, which is 
about 1 per $10^5-10^6$ years in a galaxy the size of ours
(Phinney 1991; van den Heuvel \& Lorimer 1996), then
there should be several visible structures from GRB in the interstellar
media of most spiral galaxies. Here we take a further look at the
interaction between super-supernovae and interstellar gas. 

\section{Interaction between Super-supernovae from Gamma Ray Bursts 
and Interstellar Gas}

The photons from a GRB can directly ionize and heat a large volume of
the ISM, and the fast ejecta can heat the ISM behind a shock front. At
first the GRB blastwave is relativistic, but after it slows to
sub-relativistic speeds, the subsequent interaction with the ISM will
depend mostly on the energy deposited by the ejecta. The result is a
Sedov-Taylor phase expansion, as in a supernova remnant, but with
$10-100\times$ the normal supernova energy (Wijers, Rees \& M\'esz\'aros 
1997; Waxman, Kulkarni \& Frail 1998).

The Sedov solution has radius $R$, energy $E$, pre-shock density
$\rho_0$, and time $t$ related by the equation
$R\sim\left(2Et^2/\rho_0\right)^{1/5}$.
Interior cooling follows the usual
supernova evolution. After the swept-up
shell cools, the remnant enters the pressure-driven snowplow phase
(Cioffi, McKee, \& Bertschinger 1988) at the radius
$R_{PDS}=27E_{52}^{2/7}n^{-3/7}$ pc, velocity
$v_{PDS}=490E_{52}^{1/14}n^{1/7}$ km s$^{-1}$, and time
$t_{PDS}=2.2\times10^4E_{52}^{3/14}n^{-4/7}$ yrs. Thereafter it grows as
$R/R_{PDS}\sim\left([4/3][t/t_{PDS}]-1/3\right)^{0.3}$ because of the
shell momentum and pressure from the hot cavity, until it merges
with the ambient ISM. At this time 
$t_{\rm merge}\sim 4.2\times10^6E_{52}^{0.32}n^{-0.37}v_1^{-1.43}$ years, 
the velocity has slowed to $10v_1$
km s$^{-1}$, and the radius is $R_{\rm merge}\sim140E_{52}^{0.32}
n^{-0.37}v_1^{-0.43}$ pc. Here we use the notation $E=10^{52}E_{52}$ ergs,
with preshock density $n$ in cm$^{-3}$. 
These results depend only weakly on metallicity.
The final radius may be large enough
for blowout into the halo, especially if the GRB is offset from the
midplane, but not if it is in the midplane 
with $E_{52}\sim1$, and ambient magnetic fields
confine the gas (Tomisaka 1998).

If GRB come from binary neutron star mergers, then there should be
enough events to produce giant remnants in most large galaxies.
Observational estimates from binary pulsars of the frequency of such
mergers range from $10^{-6}$ (Phinney 1991) to $8\times10^{-6}$ per year
per galaxy (van den Heuvel \& Lorimer 1996). Theoretical estimates from
stellar evolution models range from $3\times10^{-5}$ (Portegies Zwart
\& Spreeuw 1996) to $3\times10^{-4} - 3\times10^{-5}$ per year per
galaxy (Lipunov, Postnov \& Prohorov 1997a, b). Similarly, the frequency
of accretion-induced neutron star collapses in binary systems 
is estimated to be 
$\sim10^{-5}$ yr $^{-1}$ (Qin et al. 1998). When these frequencies are
multiplied by the typical remnant lifetime of several $\times10^7$ yrs,
the average number of GRB remnants per galaxy is 10 to 100.
We consider in the next section whether such remnants have
been observed already. 

\section{Giant Stellar and Gaseous Shells in Galaxies}

The Constellation III region of the Large Magellanic Cloud was the first
candidate for a super-supernova explosion (Westerlund \& Mathewson 1966;
Hodge 1967). This region has a 600-pc long stellar arc noted by these
authors, and is surrounded by a 1200-pc diameter HI ring (McGee \&
Milton 1966; Domg\"orgen et al. 1995; Kim et al. 1997) dotted with HII
regions (Meaburn 1980). There is no obvious bright stellar association
in the center that would have moved the ISM around this much (Reid et
al. 1987; Olsen et al. 1997; Braun et al. 1997). A similar stellar arc
in the galaxy NGC 6946 was attributed to super-SN explosions by Hodge
(1967), who also noted the lack of a centralized HII region (this arc is
located [106,26] mm from the lower left corner of the large image of NGC
6946 in Sandage \& Bedke [1988]). Another is in M101, at the position
(193, 165) mm from the lower left corner of the page 12 image in Sandage
\& Bedke (1988). Both the NGC 6946 and M101 features look like circular
rings of enhanced brightness with a regular outer edge and multiple arcs
of star clusters inside; they also show up as bright circular spots in
Arp's (1966) Atlas. A dozen other stellar arcs with various sizes were
attributed to large-scale explosions in galaxies by Hayward (1964), but
most of these are probably not real.

There have been many studies of supershells (Heiles
1979) without obvious central star clusters (Hu 1981; Heiles 1984).
Recent studies have found kpc-size holes and rings in irregular galaxies
(Puche et al. 1992; Radice et al. 1995) that are also devoid of obvious
centralized star formation. Radice et al. concluded that the
``supernovae hypothesis for the creation of the HI holes observed in
these galaxies is incorrect.'' Similarly Stewart et al. (1997) found in
Holmberg II that ``none of the bright FUV knots lie within the H I holes and
that they are more likely to be found immediately outside of a hole
boundary.'' Rhode et al. (1997) demonstrated for several dwarfs that
"in at least several of the holes the observed upper limits for the
remnant cluster brightness are strongly inconsistent with the SNe
hypothesis.'' 

These observations support the GRB scenario discussed in the previous
section, but the interpretation that there are no central clusters
should be viewed with some caution. We recently found (Efremov \&
Elmegreen 1998) that in the Constellation III region of the LMC, a small
cluster of 6 A-type supergiants, $\sim30$ My old, could be the remnant
of an old OB association that formed Constellation III, and that these
Constellation III stars could have caused the continued expansion of the
HI hole to make today's 1200 pc superbubble. The first cluster is barely
visible today because its brightest members have evolved off the main
sequence and dispersed.
Dwarf galaxies like the LMC generally
have little shear and a thick disk, so giant bubbles
can form slowly around old clusters and their
descendants without leaving obvious bright clusters in the center. Similar
circumstances occur in the outer spiral arms of galaxies:
shear is generally low in spiral arms, 
the outer gas disk is thick, and the outer arms have very long
flow-through times. 
Under these conditions, OB
associations and their descendants, forming and staying in the arms for
a relatively long time (50-100 My), can slowly make superbubbles without
leaving 
much evidence for star formation activity in the centers. The $\sim40$
My-old Cas-Tau association in the center of Lindblad's ring (Blaauw
1984) may be an example of such giant bubble formation -- in this case
there is shear because the solar neighborhood has emerged from the local
spiral arm already (Elmegreen 1993). Other giant bubbles are
in the southern spiral arm of M83 (Sandage \& Bedke 1988) 
and the northern arm of M51 (Block et al. 1997). The detection of 50-100 My
old clusters inside these bubbles may be difficult.

Is there other evidence for giant ISM disturbances? Rand et al. (1990)
and Dettmar (1990) found ionized loops and filaments far from the plane
in the edge-on galaxy NGC 891, and Dettmar (1992) found the same in NGC
5775. Such loops correlate with midplane star formation activity, so
they could be from normal stellar winds and supernovae (Rand et al.
1992; Dettmar 1992). 

Kamphuis, Sancisi, \& van der Hulst (1991) found a shell in M101 with a
size of 1.5 kpc and an expansion speed of $\sim50 $ km s$^{-1}$, giving
it a kinetic energy of $\sim10^{53}$ ergs; they suggested it was made by
$\sim10^3$ supernovae. Kamphuis \& Sancisi (1993) found several $10^7$
M$_\odot$ high velocity features with kinetic energies of $\sim10^{53}$
erg in NGC 6946. Vader \& Chaboyer (1995) observed a 3 kpc stellar arc
in the spiral galaxy NGC 1620, with a mass of $\sim10^7$ M$_\odot$ and a
likely expansion speed of $20-50$ km s$^{-1}$, giving it a kinetic
energy of $\sim 10^{53}$ erg; they also found an extremely bright star
cluster near the center, so they proposed the source of the expansion
was supernovae. Lee \& Irwin (1997) found four expanding HI shells at
high latitude in the edge-on galaxy NGC 3044, and estimated their masses
and kinetic energies to be $\sim10^7$ M$_\odot$ and $10^{53}-10^{54}$
erg, but because of the galaxy inclination, no central clusters could be
seen. King \& Irwin (1997) found two supershells in another edge-on
galaxy, NGC 3556, with one requiring $\sim10^{56}$ ergs of supernova
energy input according to standard models. All of these cases are
candidates for GRB explosions, but there could be old clusters in
them too, hidden by poor viewing angles, or scattered and dimmed with time
into unrecognizable forms.

Giant HI shells can also be made by high velocity cloud impacts
(Tenorio-Tagle 1981). van der Hulst \& Sancisi (1988) suggested that a
large HI complex in M101 has this origin, but Lee \& Irwin (1997) and King
\& Irwin (1997) suggested this is not the case for the giant shells they
studied because of the relative isolation of the galaxies and
lack of evidence for HI clouds around them. 
 
Evidently, there is ample evidence for kpc-size shells with masses of
$\sim10^7$ M$_\odot$, energies of $\sim10^{53}$ ergs, and no obvious
central OB associations. Some of these might be candidates for GRB
shells, but the standard explanation in terms of multiple
supernova seems acceptable too. 
Indeed, the size distribution of giant shells (Oey \& Clarke 1997)
does not reveal a clear second population that might have an origin
distinct from that of the smaller shells. 

There is a difference between 
GRB shells and shells made by multiple 
supernovae. Most of the GRB remnant expansion is the result of momentum
conservation after shell cooling and/or blowout, whereas 
the expansion around multiple supernovae
relies on continuous energy input to keep the cavity at a high
pressure (Tenorio-Tagle \& Bodenheimer 1989). 
This pressure constraint for 
the supernova model makes it difficult to build a 
shell much larger than the
disk thickness (MacLow \& McCray
1988; Tenorio-Tagle, Rozyczka \& Bodenheimer 1990). The multiple supernova
model
also has a relatively slow energy input that can be lost to radiation
inside the cavity. The average energy input rate from 1000 supernova
spread over $2\times10^7$ years inside a cavity 1 kpc in diameter is
$\sim1\times10^{-25}$ erg cm$^{-3}$ s$^{-1}$. This heating rate is
comparable to the cooling rate of $\sim1\times10^{-22}n^2$ erg cm$^{-3}$
s$^{-1}$ at $10^6$ K (Sutherland \& Dopita 1993) for normal total
interstellar pressures ($P_{\rm ISM}\sim3\times10^4$ k$_{\rm B}$).
Moreover, most star complexes still have dense cloud debris in
their vicinities, so evaporative cooling, and collisional cooling in the
high-density cloud envelopes, would remove even more supernova energy.
Thus the formation of kpc-size shells by continuous energy input from
multiple supernovae might be difficult at solar-neighborhood
or greater pressures. Combined 
GRB + supernova models for kpc-shells, or pure GRB models, might be
preferred. In a combined model, a GRB explosion 
in an aging star complex converts a
supernova-dominated $\sim500$ pc shell into a GRB-dominated 1.5
kpc shell.  

\section{Conclusions}
 
A GRB may leave a kpc-scale remnant in the interstellar medium of the
host galaxy. There is ample evidence for such disturbances in the form
of shells and high latitude filaments, 
and the number of them is consistent with the
expected frequency of GRB, but there is no definitive proof that any of
these energetic features actually required a GRB rather than 
multiple supernovae from a
star complex. If it can be shown with realistic simulations and
other studies that supernovae alone are not sufficient to make a shell
larger than $\sim1$ kpc, perhaps because of energy losses, disk blowout,
or other problems specific to the supernova model, 
then the observed kpc shells in nearby galaxies could
contain GRB remnants. 

We are grateful to the referee for useful comments.
Yu.E. appreciates the support from the Russian Foundation for the Basic
Researches, grant 97-02-17358.

\end{document}